\documentstyle[11pt,aasms4]{article}
\def\kms{\ifmmode {\rm \ km \ s^{-1}}\else $\rm km \ s^{-1}$\fi}
\def\cm{\ifmmode {\rm \ cm }\else $\rm cm$\fi}
\received{13 June 2000}
\slugcomment{Submitted to JKAS 2000}
\lefthead{Ahn, Lee, and Lee}
\righthead{Ly$\alpha$ transfer in thick and dusty medium}

\begin{document}
\title{Ly$\alpha$ Transfer in a thick, dusty, and static medium}
\author{$^1$ Sang-Hyeon Ahn, $^2$ Hee-Won Lee, $^1$ Hyung-Mok Lee}
\affil{$^1$Astronomy Program, SEES, Seoul National University, Seoul, Korea\\
$^2$ Department of Astronomy, Yonsei University}
\authoremail{sha@astro.snu.ac.kr}

\begin{abstract}
We developed a Monte Carlo code that describes the resonant Ly$\alpha$ 
line transfer in an optically thick, dusty, and static medium.
The code was tested against the analytic formula derived by
Neufeld (1990). We explain the line transfer mechanism
for a wide range of line center optical depths by
tracing histories of photons in the medium.
We find that photons escape from the medium by a series of wing scatterings,
during which polarization may develop. We applied our code to examine
the amount of dust extinction around the Ly$\alpha$ in primeval galaxies.
Brief discussions on the astrophysical application
of our work are presented.
\end{abstract}


\section{Introduction}
Recent advances of large telescopes and CCD techniques
have been unveiling very remote astrophysical objects such as
primeval star forming galaxies and damped Ly$\alpha$ galaxies.
The Ly$\alpha$ emission profiles of
these primeval galaxies are categorized into three types :
(1) pure Ly$\alpha$ emission,
(2) asymmetric or P Cyg type one, (3) broad absorption in damping wings.
Moreover, local starburst galaxies also show similar
characteristics (e.g. Kunth et al. 1998). We will not list up the
full inventory of the starburst galaxies mentioned above,
but the issue was discussed in our previous paper (Lee \& Ahn 1999).
It is plausible that dusts play an important role in the formation
of the Ly$\alpha$ line profile.
Since Ly$\alpha$ sources in a starburst galaxies are usually embeded
in optically thick and dusty media, we must investigate
the effects of dusts on the radiative transfer and line formation
of Ly$\alpha$ in the media.

Recently Meurer et al. (1999) studied the dust extinction
in local starburst galaxies, and applied the result to
the primeval galaxies in the {\it Hubble Deep Field}.
They found a tight relationship between the $\beta$ indexes
of ultraviolet (UV) continua and the
ratios of far-infrared (FIR) to UV fluxes of the galaxies.
The relationship is established
because dusts absorb Ly$\alpha$ line photons
and UV continua of $\lambda_0>912\AA$, and then re-emit in FIR.
Leitherer et al. (1999) suggested that the Ly$\alpha$
luminosity is as large as $10\%$ of the total UV luminosity
in the star forming regions.
Hence it is necessary to know what fraction of Ly$\alpha$ photons
is subject to dust extinction during the transfer
in a very thick neutral medium.

Studies of the Ly$\alpha$ line transfer in an optically thick and static 
medium has a long history.
Unno (1955) formulated the Ly$\alpha$ line transfer in a dust-free
medium, and Osterbrock (1962) proposed a simple physical
picture for understanding the resonance line transfer
in a thick medium.
Adams (1972) revised Osterbrock's picture by a Monte Carlo method
and gave a heuristic explanation on the problem.
In an analytical way, Harrington (1973) solved the problem,
for which Neufeld (1990) also gave a more general solution.
However, the studies thus far are limited to the cases where
diffusion approximation is valid and little attention has been
paid to the polarization of the Ly$\alpha$ flux emergent from those
media with anisotropic geometry and/or kinematics.
More realistic treatment of the Ly$\alpha$ line transfer 
is clearly necessary.

We have developed a sophisticated Monte Carlo code that
describes the radiative trasfer of Ly$\alpha$
resonnance photons in optically thick, dusty, and static media,
ranging from moderate optical depths to extremely large optical depths.
In Section II, we describe the basic theory including 
the numerical method of our code. The basic results are presented 
in Section III, and the possible applications of our code are 
described in Section IV.

\section{Basic Theory}
\subsection{Configuration and Review}

\begin{figure}[ttp]
  \begin{center}
    \leavevmode
    \epsfxsize = 14.0cm
    \epsfysize = 14.0cm
    \epsffile{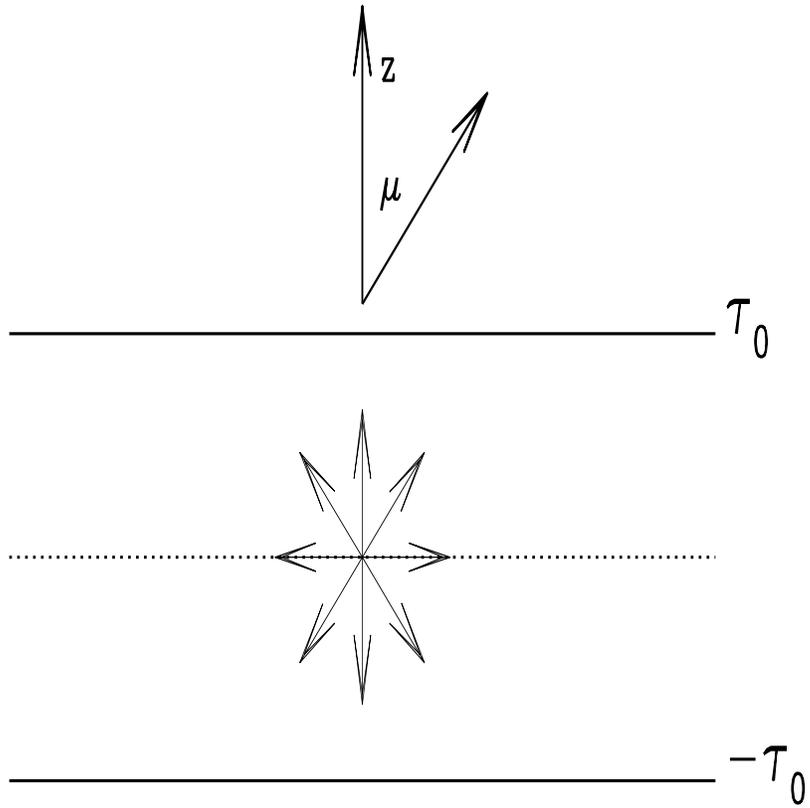}
   \end{center}
\caption{
Schematic diagram describing the configuration 
in our investigation. We adopt a plane-parallel and uniform medium 
of optical depth $\tau_0$ to both sides in $z$-direction. 
The Ly$\alpha$ photons are generated isotropically from the source 
located at the center of the medium.}
\label{nodefig}
\end{figure}

The medium used in our investigation is plane-parallel, in which hydrogen 
atoms and dust particles are mixed homogeneously.  In Fig.~1, we show the 
configuration. We consider the case that a source is located in the 
mid-plane and radiates photons isotropically. However, it is easy
to extend to other configurations.

Osterbrock (1962) proposed a simple physical
picture based on a simplified treatment of random walk processes, and 
predicted that $\langle N \rangle \propto\tau_0^2$,
where $\langle N \rangle$ is the mean number of scattering
and $\tau_0$ is the line center optical depth.
But Adams (1972) solved the problem
by a Monte Carlo method and showed that $\langle N \rangle \propto \tau_0$
rather than $\langle N \rangle \propto\tau_0^2$,
where $a$ is the Voigt parameter.
Also $\langle N \rangle$ and the peak of emergent spectra is dependent upon
$a\tau_0$ rather than $\tau_0$ alone.
He made an explanation that the line transfer process
is dominated by wing scattering, and is well approximated
by random walk processes.
He also pointed out that the transfer of Ly$\alpha$ photons
in a thick medium can be approximated as a diffusion process both in real space
and in frequency space. He suggested that in moderately thick media
(i.e. $10^3<\tau_0<10^3/a$) Ly$\alpha$ photons escape
by a single longest flight,
and for extremely thick media (i.e. $\tau_0>10^3/a$)
by a single longest excursion.

Harrington (1973) solved the radiative transfer equation presented 
by Unno (1955) in an analytic manner.
He adopted the diffusion approximation and
used the Sturm-Liouville theory to obtain the analytic solution. 
However, he considered only for one specific configuration, in which 
a monochromatic ($\lambda_0=1216\AA$) source is located 
at the center of a plane-parallel medium.
Neufeld (1990) gave analytic solutions for a more generalized version
of the problem, where the source produces arbitrary initial frequencies and
various dust extinctions were included.

In the present paper, we will consider the same problems with more general
optical depths using Monte Carlo techniques, and the results obtained
by Neufeld (1990) will be used to assure the validity of our method.
However, it is noted that our Monte Carlo approach is more faithful
to the problem by resorting to less approximation than other approaches
mentioned above.

\subsection{Optical Depth}

In the configuration stated in the previous subsection,
the optical depth is given by
\begin{eqnarray}
\tau_{\nu} (s) = \int_0^s dl \int^{\infty}_{-\infty}dv_z \ n(v_z) \sigma_{\nu}.
\end{eqnarray}

In a static medium, the Ly$\alpha$ scattering cross section is given by
\begin{eqnarray}
\sigma_{\nu} = { \pi e^2 \over m_e c }
f_{12} { \Gamma / 4\pi^2 \over (\nu - \nu_0)^2 + (\Gamma /4\pi)^2},
\end{eqnarray}
where $m_e$ is the electron mass, $c$ the light velocity,
$f_{12}=0.4126$ the oscillator strength for hydrogen Ly$\alpha$,
$\Gamma=A_{21}=3.1\times10^9 {\rm s^{-1}}$
the damping constant, and $\nu_0$ the frequency of the line center.
When we include the thermal motions of scatterers, $\nu$ is tranformed
by $\nu' = \nu - \nu_0 {v_z\over c}$, and the thermal
motion is described by the Maxwellian distribution,
\begin{eqnarray}
n(v_z) dv_z
= n_{HI} {1 \over \sqrt{\pi} v_{th}} e^{-({v_z \over v_{th}})^2} dv_z.
\end{eqnarray}
Here $v_{th}^2=2k_BT_g/m_H$, where $m_H$ is the hydrogen mass,
$k_B$ the Boltzmann constant, $n_{H}$ the neutral hydrogen density,
and $T_g$ the gas temperature.

Substituting Eqs.(2) and (3) into Eq.(1),
we obtain the expression for the optical depth,
\begin{eqnarray}
\tau_x (s) &=& { \pi^{1/2} e^2 \over m_e c } f_{12} n_{HI} s {1 \over \nu_D}
{a \over \pi}\int^{\infty}_{-\infty}du {e^{-u^2} \over (x-u)^2 + a^2} \cr
&=& 1.41\times10^{-13} T_{g4}^{-1/2} n_{HI} s H(x,a),
\end{eqnarray}
where $T_{g4}\equiv {T_g / 10^4{\rm\ K}}$,
\begin{eqnarray}
\nu_D = {v_{th} \over c} \nu_0,
\end{eqnarray}
\begin{eqnarray}
x\equiv{\Delta\nu \over \Delta\nu_D}\equiv{\nu-\nu_0 \over \Delta\nu_D},
\end{eqnarray}
\begin{eqnarray}
u\equiv {v_z \nu_0 \over c \Delta\nu_d},
\end{eqnarray}
and
\begin{eqnarray}
a\equiv {\Gamma \over 4 \pi \Delta\nu_d}.
\end{eqnarray}
Here $H(x,a)$ is the Hjerting function or the H-function,
and the specific values in Rybicki \& Lightman (1979) are used.
Also the Voigt parameter $a$ is given by
\begin{eqnarray}
a = 4.7\times 10^{-4} T_{g4}^{-1/2}.
\end{eqnarray}
Therefore, the total optical depth for a given system is
\begin{eqnarray}
\tau_x = 1.41\times10^{-13} T_{g4}^{-1/2} N_{HI} H(x,a)
= \tau_0 H(x,a),
\end{eqnarray}
where the column density of neutral hydrogen, $N_{HI}=n_{HI}L$
with $L$ being the physical size of the slab, and $\tau_0 \equiv
1.41\times10^{-13} T_{g4}^{-1/2} N_{HI}$ is the line center optical depth.

\subsection{Monte Carlo Code}

In this subsection, we present a detailed description of
the Monte Carlo procedure.
There are a few Monte Carlo approaches to the resonance
line transfer in an optically thick and static medium in
the literature (e.g. Adams 1972, Harrington 1973, Gould \& Weinberg 1996).

The Monte Carlo code begins with the choice of the frequency
and the propagation direction $\vec k_i$ of an incident photon
from an assumed Ly$\alpha$ profile, which is assumed to
be monochromatic in our study.

Then we determine the next scattering site
separated from the initial point by the propagation length defined by
\begin{eqnarray}
S \equiv \tau_0 s 
= { \tau \over {H(x,a) + \tau_d/\tau_0}} \simeq {\tau \over H(x,a)},
\end{eqnarray}
where the optical depth is assumed to be composed of
hydrogen ($\tau_0 H(x,a)$) and dust ($\tau_d$) parts.
Here, $\tau_d$ is the dust optical depth which is treated as a constant
for the emission line and whose Galactic value is given
by Draine \& Lee (1984),
and $\tau=-\ln ({\it \bf R})$, where ${\bf R}$
is a uniform random number in the interval $(0,1)$
generated by a subroutine $ran2()$ suggested by Press et al.(1989).

The emitted photon traverses a distance $S$,
and is scattered off by hydrogen atoms until $|z|>\tau_0$ for a slab geometry.
In this scattering event
the frequencies of the absorbed photon and the re-emitted one
in the rest frame of the scatterer should be matched.

In contrast to the case of a thin medium,
for a very thick medium
the scattering in the damping wings is not negligible.
A careful treatment needs to be exercised to distinguish the scattering
in the damping wings from the resonance scattering,
because they show quite different behaviors in the properties
including the scattering phase function and the polarization
(Lee \& Blandford 1997).

Because the natural line width is much smaller than the Doppler width,
the local velocity of the scatterers that can resonantly
scatter the incident photon is practically a single value.
However, when the scattering occurs in the damping wings,
the local velocity of the scatterer may run a rather large range.
Therefore, in order to enhance the efficiency of the Monte Carlo method,
it is desirable to determine the scattering type before we
determine the local velocity $\vec u$ of the scatterer.

We present a more quantitative argument about the preceding remarks.
Under the condition that a given photon is scattered
by an atom located at a position $s$, the local velocity component
$u$ along the direction $\vec k_i$ is chosen from the 
normalized distribution,
\begin{eqnarray}
f(u) = {e^{-u^2} \over (x-u)^2 + a^2 }
\left[{\pi \over a} H(x,a)\right]^{-1}.
\end{eqnarray}
Here, the Hjerting function or the Voigt function $H(x,a)$ 
is evaluated by a series expansion in $a$,
i.e.,
\begin{eqnarray}
H(x,a) &=& H_0(x,a) + a H_1(x,a) +a^2 H_2(x,a) \nonumber \\
&+& a^3 H_3(x,a) + \cdots,
\end{eqnarray}
where $H_n(x,a),~{n=0,1,2,3}$ are tabulated by Gray (1992).

Because of the smallness of $a$, the function $f$ has
a sharp peak around $u\approx x$, for which the scattering is resonant.
Therefore, the probability $P_r$ that a given scattering is resonant
is approximately given by
\begin{eqnarray}
P_r  &\simeq &\int_{-\infty}^{\infty} d(\Delta u)
{e^{-x^2} \over  (\Delta u)^2 + a^2 }\left[{\pi \over a} H(x,a)\right]^{-1}
\nonumber \\
& =& {e^{-x^2} \over H(x,a)}.
\end{eqnarray}
The probability that scattering occurs in the damping wings is
\begin{eqnarray}
P_{nr} = 1 - P_r.
\end{eqnarray}

In the code we determine the scattering type
in accordance with the scattering type probabilities $P_r$ and $P_{nr}$.
If scattering is chosen to be resonant, then we set $u=x$.
Otherwise, the scattering occurs in the damping wings, and $u$ is chosen
in accordance with the velocity probability distribution given by Eq.(12).

We give the propagation direction $\vec k_f$ of a scattered photon
in accordance with the phase function completely faithful to the
atomic physics (Lee \& Ahn 1998).
The scattered velocity component $v_\perp$
perpendicular to the initial direction $\vec k_i$
on the plane spanned by $\vec k_i$ and $\vec k_f$ is also governed by
the Maxwell-Boltzmann velocity distribution, which is numerically obtained
using the subroutine ${\it gasdev}()$ suggested by Press et al. (1989).
The contribution $\Delta x$ of the perpendicular velocity component
$v_\perp$ to the
frequency shift along the direction of $\vec k_f$ is obviously
\begin{eqnarray}
\Delta x = v_\perp [1-(\vec k_i \cdot \vec k_f )^2]^{1/2}/c.
\end{eqnarray}

Therefore, the frequency shift $x_f$ of the scattered photon is given by
\begin{eqnarray}
x_f = x_i - u + u(\vec k_i \cdot {\vec k_f} )
    + v_\perp [1-(\vec k_i \cdot \vec k_f )^2]^{1/2},
\end{eqnarray}
where $x_i$ is the frequency shift of the incident photon.

In each scattering event
the position of scattered photon is checked, and
its path length is added.
If the photon escapes from the medium or $|z|<\tau_0$,
we collect that photon according to
its frequency and escaping direction.
In collecting photons, we add the weighted fraction
considering the dust extinction corresponding to the
total path length of the photon.
The whole procedure is repeated until typically
about $10^3$ photons in each frequency bin are collected.

\section{Results}

\subsection{Tracing the Scattering Processes}

\begin{figure}[ttp]
   \begin{center}
    \leavevmode
    \epsfxsize = 11.0cm
    \epsfysize = 11.0cm
    \epsffile{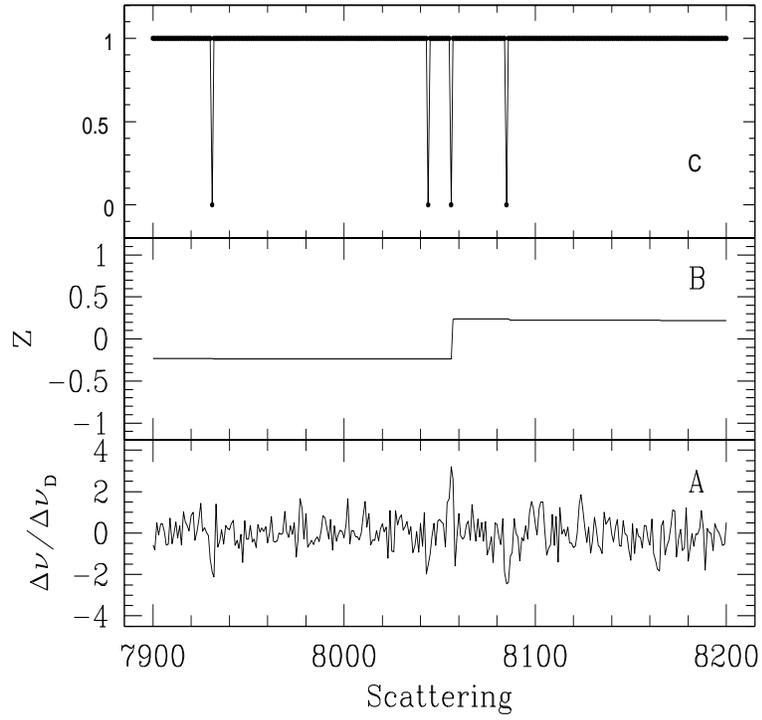}
    \end{center}
\label{nodefig}
\caption{
History of a photon. $x$-axis is the scattering number,
and we show frequencies of the photon in Panel A, its locations in
panel B, and its type of scattering in panel C. In panel C, `0' means
wing scattering, and `1' means core scattering.}
\end{figure}

\begin{figure}[ttp]
   \begin{center}
    \leavevmode
    \epsfxsize = 11.0cm
    \epsfysize = 11.0cm
    \epsffile{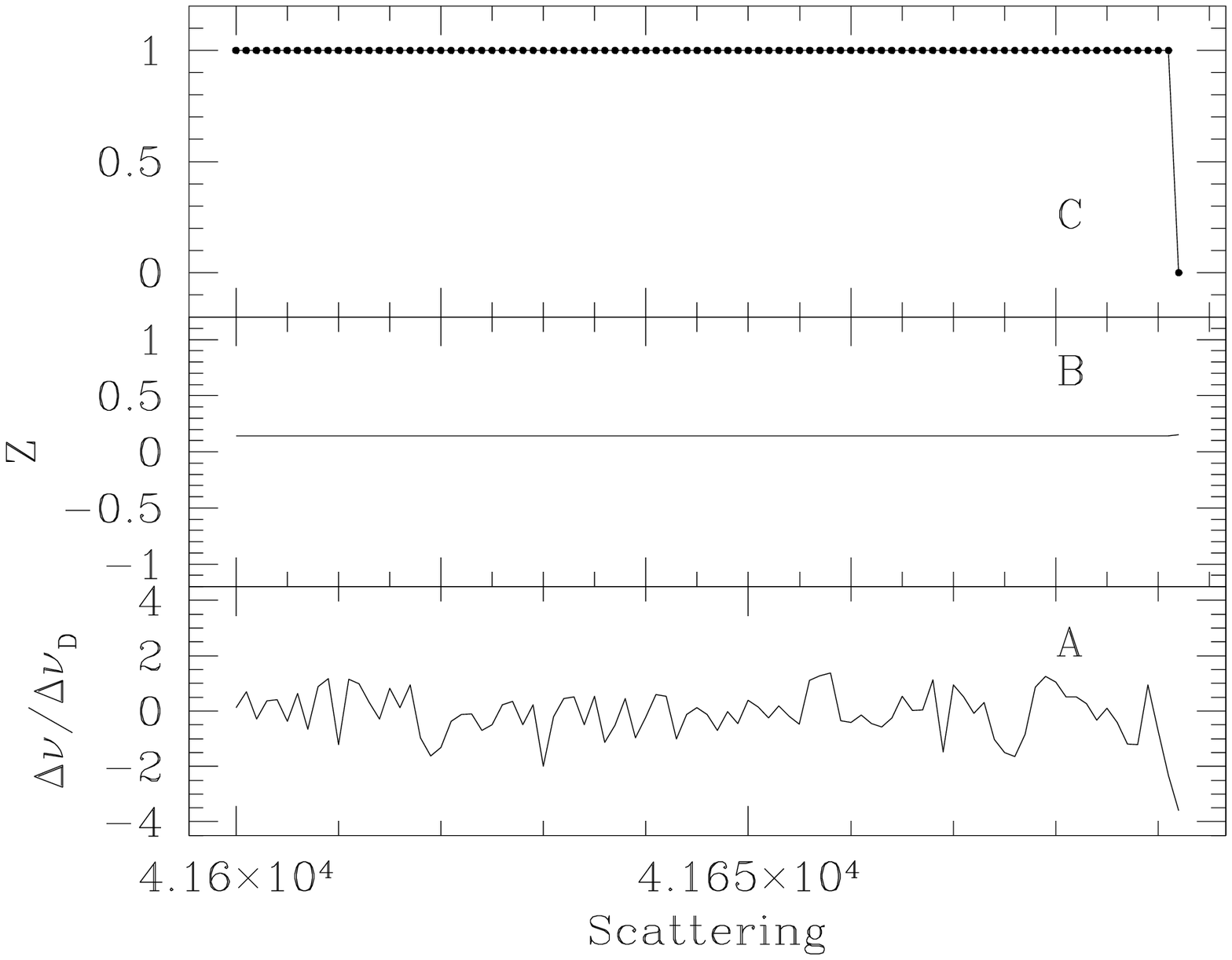}
    \end{center}
\label{nodefig}
\caption{
History of a photon just before escape.
Panels represent the same quantities to those in Fig.~2.
We can see that the photon experiences wing scattering
just before escape.}
\end{figure}

Adams (1972) gave a physical picture that describes the Ly$\alpha$ line
transfer in extremely thick media of $\tau_0>10^3/a$,
where $a=4.71\times10^{-4}$ for $T_g=10^4{\rm\ K}$.

Initially a core photon becomes a wing photon
after a sufficient number of scatterings and encountering with a 
violently moving atom.  Due to the small optical depths,
wing photons traverse much further in physical space than core photons.
When $\tau_0<10^3$, a photon can
directly escape from a medium once it becomes a wing photon.
This process is called `a single longest flight' by Adams (1972).

As the line center optical depth gets larger
to be an intermediate optical depth, $10^3<\tau_0<10^3/a$,
wing photons cannot directly escape from the medium,
but experience a large number of core scatterings 
followed by a few wing scatterings.
However, after these wing scatterings, photons may
return back to the core because of the so called `the restoring force'
(Osterbrock 1962).

For the case of the extremely thick medium we can describe this process 
using a diffusion approximation,
where the $rms$ frequency shift per scattering is the thermal
Doppler width ($v_{th}$) or $x=1$,
and the mean frequency shift per scattering is $\Delta x = - 1/|x|$ that
plays the role of the restoring force. This restoring force is caused
by relatively large probability that photons scatter in the core.
However, for a medium with an intermediate optical depth,
the first variance of frequency is not quite large
because the optical depth is not enough
to enlarge the probability for wing scattering at frequencies
far from the line center.
Hence, photons often experience a small number of single longest flights 
before escape, and we may call these processes as `random wandering'.
Therefore, in this particular case, we think that a simple diffusion
approximation is not adequate and that a more accurate approach such as
a Monte Carlo method is needed.

For a medium of an even higher optical depth, $\tau_0>10^3/a$,
a series of wing scatterings occur when wing photons are
generated. In this case the wandering occurs both in real space and in 
frequency space. This process is called `an excursion,' and photons escape
the medium by `a single longest excursion' (Adams 1972).

These processes can be seen in our Monte Carlo results. 
In Fig.~2 we have shown the trajectory of a photon that transfers 
in a medium with $\tau_0=10^4$.
At the 8055th scattering, the photon is scattered off
by an atom that moves fast and became a wing photon.
At the same time, the photon traverses a longer distance
in physical space, and also wing scattering happens.

Before escaping from the medium, the photon experiences a series of
wing scatterings. In Fig.~3 we show the history of the same
photon, and here we can see that the last scattering occurs in the wing.
We checked that the larger $\tau_0$, the larger the number of last wing 
scatterings, which may be regarded as `a single longest excursion.'
These last wing scatterings possibly induce polarization
because wing scattering has a Rayleigh phase function (Stenflo 1980).
We have conducted an investigation on this topic,
and the results will be published in another paper.

\subsection{Emergent Ly$\alpha$ Profile for Dust Free Media}

\begin{figure}[ttp]
   \begin{center}
    \leavevmode
    \epsfxsize = 10cm
    \epsfysize = 10cm
    \epsffile{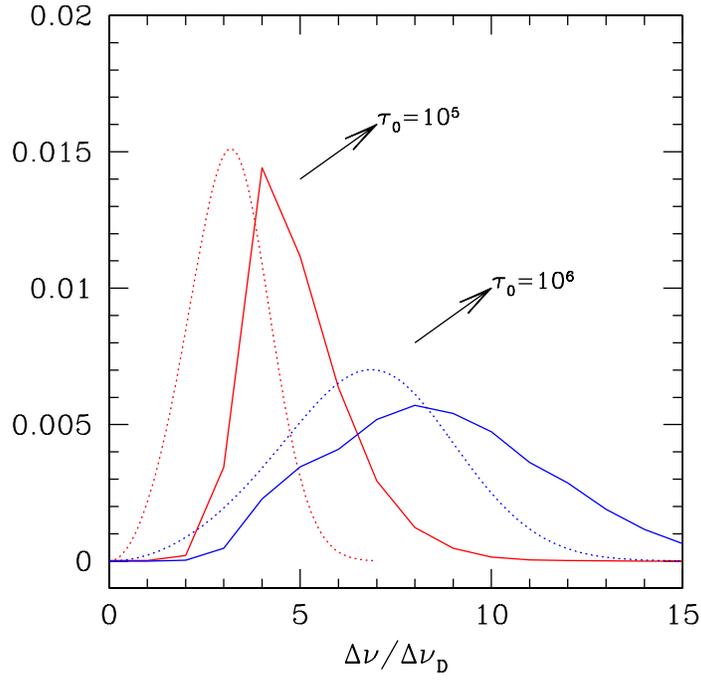}
    \end{center}
\label{nodefig}
\caption{
Our emergent profiles (solid lines) are 
compared with Neufeld's analytic solution (dotted lines).
$x$ axis is a frequency in units of the thermal width, and the total
flux of the line is normalized to $1/8\pi$ in accordance with
Neufeld's normalization.  The profiles are symmertic about the origin, 
$\Delta \nu/\Delta\nu_D=0$.}
\end{figure}

Neufeld (1990) derived an analytic solution, for $(a\tau_0)^{1/3} \gg1$.
According to Unno (1955), Harrington (1973), and Neufeld (1990),
for the cases of dust-free media,
the Ly$\alpha$ line transfer equation is given by
\begin{eqnarray}
{\partial^2 J \over \partial\tau^2} + {\partial^2 J \over \partial\sigma^2}
= -3 \phi(x) {E(\tau,x) \over 4 \pi},
\end{eqnarray}
where
\begin{eqnarray}
\sigma = \left( {2 \over 3} \right)^{1/2} \int_0^x {dx' \over \phi(x')},
\end{eqnarray}
and $\phi(x)$ is the normalized Voigt function given by
\begin{eqnarray}
\phi(x) = {1 \over \sqrt{\pi}} H(x,a).
\end{eqnarray}
Here the boundary conditions are
\begin{eqnarray}
J(\tau,\pm\infty) = 0
\end{eqnarray}
\begin{eqnarray}
J(\pm \tau_0,x) = \pm 2 H(\pm\tau,x) = \mp {2 \over 3\phi(x)}
\left( {\partial J \over \partial\tau} \right)_{\pm\tau_0}.
\end{eqnarray}

Since the source function $E(\tau,x)$ in the right hand side of Eq. (19)
is very sharply peaked, we can apply the Green function method to the problem.
With an approximaton 
\begin{equation}
\phi(x)\simeq a/\pi x^2
\end{equation}
and an application of the Sturm-Lioville theory, Neufeld (1990)
derived a solution for the case of a mid-plane source,
\begin{eqnarray}
J(\pm\tau_0,x) = {\sqrt{6} \over 24}{x^2 \over a\tau_0}
{ 1 \over \cosh[(\pi^4/54)^{1/2}(|x^3-x_i^3|/a\tau_0)]}.
\end{eqnarray}

We perform Monte Carlo calculations for a monochromatic source with
$x_i=0$, and compare the result with Eq.(24). Fig.~4 shows
our results for $\tau=10^5$ and $\tau_0=10^6$.
The solid lines are the results of Monte Carlo calculation,
and the dotted lines are those obtained by Neufeld.
The shapes of the profiles agree with each other,
and we emphasize that our Monte Carlo code is written to incorporate
all the quantum mechanics associated with 
both resonant and non-resonant Ly$\alpha$ scattering.

The figure shows an overall disagreement: 
the Monte Carlo solution appears to be translated from the Neufeld's
solution by an amount of $x\simeq 3$, which corresponds to the
frequency shift where the wing scattering becomes important and
the profile function $\phi(x)$ can be nicely approximated by
Eq.~(24). In fact, in the cases of dust free media, 
no loss of Ly$\alpha$ line photons is permitted, which guarantees
the flux conservation. 
The profile function $\phi(x)$ can not be approximated by Eq.(23)
at $x\le 3$, and therefore, Neufeld's calculation underestimates 
the amount of core photons that are removed and ultimately redistributed 
to the wing regimes, 
which reduces the amount of diffusively trasferred wing photons.

\subsection{Survival Fraction for Dusty Media}

\begin{figure}[ttp]
  \begin{center}
    \leavevmode
    \epsfxsize = 10cm
    \epsfysize = 10cm
    \epsffile{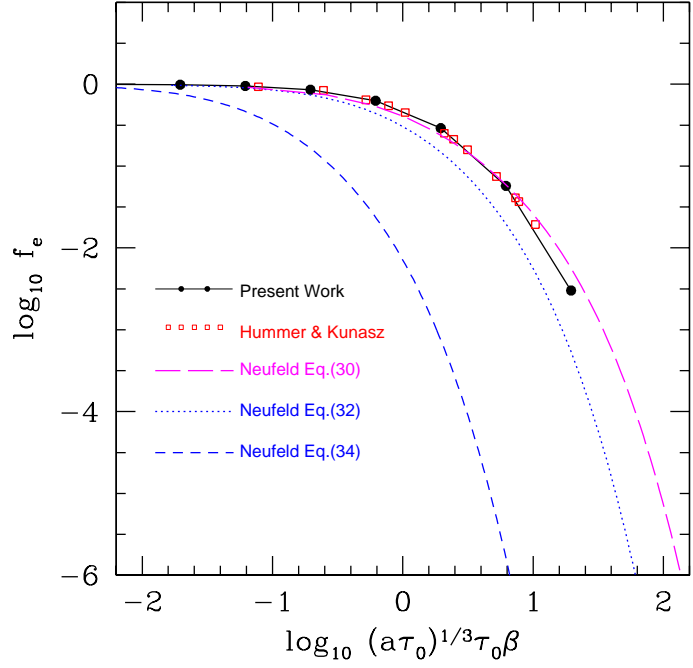}
   \end{center}
\label{nodefig}
\caption{
Survival fraction of photons in our Monte Carlo 
calculations for dusty media of $\tau_0=10^6$ is compared with 
previous works. 
Long dashed line stands for
the case of a variable effective frequency, and dotted line 
for the case of a constant effective frequency. Short dashed line represents
Neufeld's derivation for $Y_0$ by considering average scattering number
of photons in media with intermediate optical depth.}
\end{figure}

Neufeld (1990) also provided an analytic solution for the problem of
the radiative transfer in a dusty medium.
Here we will concentrate on the continuum absorption case.
Harrington (1973) and Neufeld (1990) formulated this problem by
\begin{eqnarray}
{\partial^2 J \over \partial\tau^2} + {\partial^2 J \over \partial\sigma^2}
= -3 [\phi(x)+\beta_t(x)] \left[ {E(\tau,x) \over 4 \pi} - \beta(x) J \right],
\end{eqnarray}
where
\begin{eqnarray}
\sigma = \left( {2 \over 3} \right)^{1/2} \int_0^x
{dx' \over \phi(x')+\beta_t (x^{'})}.
\end{eqnarray}
We assume that the total dust opacity $\beta_t(x)\approx 0$
becasue $\phi(x)\gg\beta_t$.
Here, $\beta$ is an absorptive part of dust opacity per atomic hydrogen,
which is given by
\begin{eqnarray}
\beta = 1.0\times 10^{-8} T_{g4}^{1/2} {n_H \over n_{HI}} \xi_d.
\end{eqnarray}
Here
\begin{eqnarray}
\xi_d
=(1-A) {\tau_d \over N_{HI}} / \left[(1-A) {\tau_d \over N_{HI}}\right]_G,
\end{eqnarray}
where A is albedo and the subscript $G$ represents the Galactic value.
A dust opacity per a hydrogen atom is given by
\begin{eqnarray}
{\tau_d \over N_{HI}} = f_d (\sigma_{sca} + \sigma_{abs}),
\end{eqnarray}
where $\tau_d$ is dust optical depth, $f_d$ is dust-to-gas ratio,
$\sigma_{sca}$ and $\sigma_{abs}$ are the scattering
and the absorption cross sections, respectively.
According to Draine and Lee (1984), the mean value of the dust
optical depth for Our Galaxy is given by
$\tau_d / N_{HI} = 1.6 \times 10^{-21}$ at $\lambda_0=1216\AA$.

In our Monte Carlo code,
we give $\tau_d$, $A$, as well as $N_{HI}$ or $\tau_0$.
Therefore we can compare these two kinds of calculations.
Neufeld also gave solutions for two kinds of
approximation: (1) variable effective scattering frequency,
(2) constant effective scattering frequency.
In the former approximation, the survival fraction of photons
is given by
\begin{eqnarray}
f_e = \left[ 1 + \sum_{k=1}^\infty \left( {y_0^{2k} \over k!}
{ 1 \over 3 \cdot 7\cdot 11\cdot \cdot \cdot \cdot(4k-1)} \right) \right]^{-1},
\end{eqnarray}
where $y_0 \equiv {3 \over 2} ({16 \over \pi})^{1/6} a^{1/6} \beta^{1/2}
|\tau_0|^{2/3}$.

In the latter approximation, the total fraction of photons which
escape the slab is given by
\begin{eqnarray}
f_e = { 1 \over \cosh (Y_0)},
\end{eqnarray}
where
\begin{eqnarray}
Y_0 = \left[ {3(a\tau_0)^{1/3}\beta\tau_0 \over \pi \zeta^2} \right]^{1/2},
\end{eqnarray}
and $\zeta$ is determined to be $\zeta=0.525$
by fitting the results of Hummer and Kunasz (1980)
to the following relation between peak frequency ($x_s$) and the $a\tau_0$,
\begin{eqnarray}
x_s=\zeta (a\tau_0)^{1/3}.
\end{eqnarray}
%Neufeld determined $\zeta=0.525$, and so
%\begin{eqnarray}
%Y_0 = \left[ {3.46(a\tau_0)^{1/3}\beta\tau_0} \right]^{1/2}.
%\end{eqnarray}

The other way of determining $Y_0$ is provided by Neufeld
(see Eq.~4.35 in his paper), who considered the mean scattering length
in the result of Hummer and Kunasz (1980).
The consideration leads to an survival fraction $1/\cosh(Y_0)$ with
\begin{eqnarray}
Y_0 = \left( 31.8\beta\tau_0 \right)^{1/2}.
\end{eqnarray}
He argued that this relationship may be applicable even in the
range of intermediate optical depth, $10^3<\tau_0<10^3/a$.

In Fig.~5 we show the main result on the survival fraction
of Ly$\alpha$ photons in a dusty medium with $\tau_0=10^6$ and $T=10^4{\rm\ K}$.
Our results are in good agreement with Neufeld's approximate solutions
and those of Hummer and Kunasz. Our results are also consistent with
Neufeld's approximation solution for the approximation of
a constant effective frequency.

On the other hand, although Neufeld noted that Eq.(34) above is valid
even in intermediate optical depths, according to our results,
this statement is not true.

%%%%%%%%%%%%%%%%%%%%%%%%%%%%%%%%%%%%%%%%%%%%%%%%%%%%%%%%%%%%
\section{Discussion}
In this paper we described the Ly$\alpha$ line transfer in
an optically thick, dusty, and static medium.
We took accurate atomic physical considerations in
the Monte Carlo code, and examined the effect of dusts
on Ly$\alpha$ line formation in optically thick media.

For dust free media we confirmed the line transfer mechanism,
in which a number of longest flights occur.
We emphasize that photons experience a series of wing scatterings
at the moment of escape, at which stage significant polarization
can be developed. 
We also confirmed that line profiles from our Monte Carlo calculations
agree well with the analytic solution derived by Neufeld (1990).

For dusty media, we calculated the survival fraction of
photons for $\tau_0=10^6$ and $T=10^4K$,
and found that our results are in good agreement with 
those obtained by Hummer and Kunasz (1980) and with those 
of Neufeld's approximation solution.

The computational speed is limited
for extremely thick media with $a\tau_0>10^3$.
This difficulty can be overcome by considering
only wing scatterings (Adams 1972).
Hence, we are now writing
a Monte Carlo code similar to that of Adams.
Also we will investigate effects of following various
deviations from our simplified configuration:
the initial line width, the bulk motion of scatterers,
the geometrical shape of scattering media, the degree of
homogeneity of the dusts, and the distribution form of photon sources.

One of main goals of our research is to develop a tool
to estimate the dust abundance in accordance with
the usual method using the Balmer decrement (see Osterbrock 1989)
and the slope of UV continuum (Fall \& Pei 1989).
This may help us correct properly the dust extinction
of emission lines in starburst galaxies.
Especially this will enable us to estimate the star formation
rate of those galaxies which are located near or far.

The other ramification is the formation of the damped Ly$\alpha$
absorption (DLA) in many quasar spectra.
The dust contents in DLA galaxies may deform
the absorption profile. This may spoil the usual
Voigt profile fitting procedure, and subsequent
derivation of physical values can be erroneous.

\acknowledgments
HML and SHA are grateful for the financial support of Brain Korea 21 project.
HML was also supported by KOSEF grant no. 1999-2-113-001-5.
HWL gratefully acknowledges support from the BK21 project.

%%%%%%%%%%%%%%%%%%%%%%%%%%%%%%%%%%%%%%%%%%%%%%%%%
\end{document}